\documentclass[doublespacing]{elsart}
\usepackage{epsfig}

\begin{document}

\center

\maketitle
\begin{frontmatter}
\title{Radioactive Krypton Background Evaluation Using Atom Counting}

\thanks[corr]{Corresponding author D.N. McKinsey. Email:
daniel.mckinsey@yale.edu}
\author[yu]{D.\,N. McKinsey} and
\author[uc]{C. Orzel}
\address[yu]{Yale University, New Haven, CT
06511, USA}
\address[uc]{Union College, Schenectady, NY 12308, USA}

\begin{abstract}

The beta-decay of $\rm^{85}Kr$ is a significant radioactive background
for experiments that use liquified noble gases to search for dark
matter and measure the low-energy solar neutrino flux.  While there
are several proposed methods for reducing Kr levels in these
experiments, an independent technique is needed for measuring very low
Kr levels.  By selectively exciting Kr atoms to a metastable state,
capturing them in a magneto-optical trap (MOT), and detecting
fluorescence from the trapped atoms, individual Kr atoms can be
counted with a high signal-to-noise ratio.  This approach could be
used to ascertain Kr impurity levels in other noble gases, with an
estimated sensitivity of $3\times 10^{-14}$.

\end{abstract}
\end{frontmatter}

\section{Introduction}

For a variety of ongoing and proposed experiments, the beta-decay of
$\rm^{85}Kr$ is a significant radioactive background.  With a current
isotopic abundance of $\rm 2.5 \times 10^{-11}$ in natural
Kr\cite{Loosli00} and a half-life of 10.76 years, $\rm^{85}Kr$
gives a beta-decay activity of $3.7 \times 10^{5}$ Bq per kg of
natural Kr.  In the Earth's atmosphere, with a Kr content of 1.1 ppm,
this corresponds to a $\rm^{85}Kr$ beta decay rate of $\rm
1.5\,Bq\,m^{-3}$.  The beta decay of $\rm^{85}Kr$ extends up to 687
keV, and so Kr impurities are a concern for experiments that
are searching for rare events at low energies.

Radioactive background from $\rm^{85}Kr$ beta decay is a prime concern
for experiments that use liquid Xe as a medium for the direct
detection of dark matter in the form of Weakly Interacting Massive
Particles (WIMPs).  In general, experiments designed for direct WIMP
searches need to have high sensitivity to nuclear recoils as well as
low intrinsic radioactivity.  For a more detailed description of
WIMPs, their theoretical motivation, and the basic approach to their
detection, see Ref.~\cite{Jun96}.  Liquid Xe is a promising medium for
WIMP detection, with a large nucleus, high density, large
scintillation yield, and high charge drift velocity.

Several experiments exist or are underway that use liquid Xe as a pure
scintillation\cite{Aki02,Ber01} detector or employ a combination of
prompt scintillation and proportional
scintillation\cite{Cli02,Smi03,Apr02,Mor03,Suz03}.  However, since Kr
is typically present in research-grade Xe gas at approximately 10 ppm,
steps must be taken to remove Kr and its attendant $\rm ^{85}Kr$
decay.  At present, Xe can be purchased commercially with a guaranteed
Kr level of not more than 20 ppb, but because of the limited
sensitivity of conventional measurement of Kr impurities, lower limits
cannot be guaranteed.  Both the production of Xe with low levels of Kr
and the measurement of low levels of Kr in Xe are necessary to
evaluate the radioactive background, an essential step for the
continued success of Xe as a WIMP detection medium.  For example, the
XENON detector requires natural Kr contamination to be 150 ppt or
less\cite{Tom}.

Liquid Xe has also been proposed as a detection medium for
measuring the $p-p$ solar neutrino flux\cite{XMASS}, and in this
case the requirement for Kr contamination is even more stringent:
Kr/Xe $\le$ $ 4 \times 10^{-15}$.  This is similar to the
requirement for the CLEAN detector\cite{McK00,Mic02,Hor02,McK04},
which will use liquid Ne for solar neutrino and WIMP detection. In
CLEAN, Kr can be easily removed from Ne by selective adsorption on
cold charcoal, but, as in the Xe dark matter detectors, it would
be highly useful to have an independent test for Kr contamination.
For more information on the scientific motivation for detecting
low energy solar neutrinos, see Ref.~\cite{Bah03}.

Low energy solar experiments using a liquid organic scintillator,
such as the Borexino\cite{Borexino,Arp02} and
KamLAND\cite{KamLAND} experiments (designed to detect solar
neutrinos emitted in the $\rm ^{7}Be + e^{-} \rightarrow\, ^{8}B +
\nu_{e}$ reaction) also require very low Kr levels because $\rm
^{85}Kr$ decay creates events that directly interfere with the
signal of interest.  However, it is expected that these
experiments can be purged of dissolved air (with its attendant Kr
content) with pure gaseous nitrogen from liquid nitrogen boiloff.
In addition, $\rm Kr/N_{2}$ can be determined by removing the $\rm
N_{2}$ with a getter, then analyzing the remaining noble gas
contaminants using a standard mass spectrometer\cite{Heu03}.

For Kr contamination in other noble gases such as Ne and Xe, which
are promising for low background measurements, it is not practical
to separate the Kr chemically.  Instead, a possible approach to
the measurement of very small Kr quantities in Ne and Xe is the
introduction of these gases into a system that can trap and count
individual Kr atoms with high efficiency, while simultaneously
maintaining a high throughput.  The technique of Atomic Trap Trace
Analysis (ATTA), recently developed by Z.\,T.\,Lu and
collaborators\cite{Kr}, could be used to measure Kr impurities
with a sensitivity that is not possible using other currently
available methods. This paper evaluates the possibility of using
ATTA for measuring trace quantities of Kr in Ne or Xe gases.

The ATTA approach could also be used to measure Ar contamination
in Ne or Xe gas, again with the goal of evaluating radioactive
background.  The isotope of concern in this case is $\rm ^{39}Ar$,
which is produced in the atmosphere by $(n,2n)$ reactions, has a
half-life of 269 years, and has a beta decay endpoint of 565 keV.
It is present in the atmosphere at $\rm ^{39}Ar/Ar$ = $(8.1 \pm
0.3) \times 10^{-16}$, as shown by direct low-level
counting\cite{Loo68,Loo83} and by accelerator mass
spectroscopy\cite{Ar}. We note that this experimental value
conflicts with the theoretical treatment presented by the ICARUS
group\cite{Cen95}.  Assuming that the experimental values are
correct, $\rm^{39}Ar$ gives a beta-decay activity of 1.0 Bq per kg
of natural Ar.  In the Earth's atmosphere, with an Ar content of
0.93\%, this corresponds to a $\rm^{39}Ar$ beta decay rate of $\rm
1.6 \times 10^{-2}\,Bq\,m^{-3}$ in air.  Though a given quantity
of Ar contamination results in much less radioactive background
compared to the same quantity of Kr contamination (for example,
the CLEAN experiment\cite{McK04} will require Ar/Ne $\le
10^{-10}$), it is useful to be able to measure Ar contamination
independently.  The atom trapping and counting scheme described in
this paper could equally be applied to the assay of trace
quantities of Ar. Indeed, the transition wavelengths and cooling
parameters for Ar and Kr are similar enough that the same
apparatus can be used for both.

\section{Experimental Approach}

Laser cooling and atom trapping methods rely on light pressure
forces from photon scattering to reduce the average atomic
velocity in a vapor, thus cooling the atoms to microkelvin
temperatures\cite{review}.  This process offers exceptional
isotopic selectivity, as the Doppler shifts due to the motion of
the cold atoms are many times smaller than the isotope shifts of
the atomic resonance frequencies.  Laser cooling has been used to
produce ultra-cold samples of roughly $20$ different elements,
including rare and unstable isotopes (\cite{Kr,Fr,Na,Rb}). The
ability to selectively manipulate extremely rare isotopes holds
great promise for environmental monitoring and geological
dating\cite{Kr,Sturchio,Moore}, and the ATTA technique can be
extended to measure background impurities in rare-gas samples,
which are an important source of systematic error in new neutrino
detection experiments. Since the $\rm^{85}Kr/Kr$ isotopic ratio is
well known, a limit on natural Kr derived using such a system
would imply a proportionally smaller limit on $\rm ^{85}Kr$.

\begin{figure}[!hbt]
\begin{center}
\epsfig{file=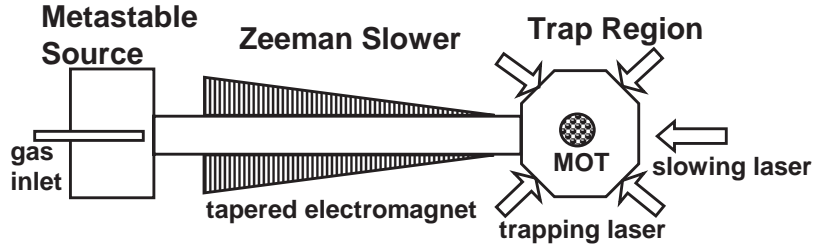, width=12cm}
\end{center}
\caption{Schematic of the proposed experimental apparatus for
measuring krypton impurities.}
\label{fig:ATTA_apparatus}
\end{figure}

The apparatus used for ATTA (Fig.~\ref{fig:ATTA_apparatus})
consists of a magneto-optical trap (MOT) loaded from a slowed
atomic beam. An atom source produces a beam of atoms moving at
thermal velocities (for Kr, $v_{th} \sim 300$\,m/s for $T \sim
300$\,K). These atoms are decelerated by the light force from a
counter-propagating laser, and pass through a spatially varying
magnetic field that produces a Zeeman shift to compensate the
changing Doppler shift as the atoms slow from $300$\,m/s to $\sim
1$\,m/s\cite{Zeeman}. Slow atoms from the beam are then captured
in the MOT, where a combination of lasers and inhomogeneous
magnetic fields trap and further cool the atoms\cite{MOT}. A
typical MOT loaded from an atomic beam will confine $10^7$ atoms
at a temperature of $\sim 100$\,$\mu$K.

The ATTA technique measures isotopic abundances by counting
trapped atoms to determine the trap loading rates for an isotope
of known abundance and a rare isotope of unknown abundance. The
loading rate of the rare isotope is then compared to the loading
rate for the more common isotope (measured separately). Assuming
the same detection efficiency for the two isotopes, the ratio of
loading rates gives the relative abundance. The abundances of
$^{81}$Kr ($1.0\times 10^{-12}$) and $^{85}$Kr ($1.5 \times
10^{-11}$) have been measured by comparing their loading rates to
that of $^{83}$Kr\cite{Kr}.

Measuring krypton impurities in samples of other rare gases
requires measuring the loading rate for $^{84}$Kr in a sample of
unknown abundance, and comparing to the loading rate from other
samples of known abundance. Conventional measurement techniques
(mass spectrometry, gas chromatography) can be used to measure Kr
impurities at the ppm level. Samples with known Kr abundance can
be prepared, and loading rates measured to serve as a reference
for comparison with higher purity samples.

The ATTA apparatus is a standard laser cooling system, with two
exceptions: In order to trap and cool krypton atoms, we must
prepare them in a metastable state, requiring a more complicated
atom source than for other atomic species; and in order to detect
rare isotopes, we must have a detection system capable of
detecting a single atom in the MOT region.

\subsection{Metastable Atom Source}

Laser cooling of rare-gas atoms in the ground state is
impractical, as the transition wavelengths needed to excite atoms
from the ground state to any excited states are prohibitively
short ($\leq 126$\,nm for Kr) given current laser technology. The
lowest excited states of these atoms, however, have extremely long
lifetimes (tens of seconds), and serve as an effective ground
state for laser cooling. Laser cooling and trapping of atoms in
these metastable states has been demonstrated in all of the stable
rare-gas species.

\begin{figure}[!hbt]
\begin{center}
\epsfig{file=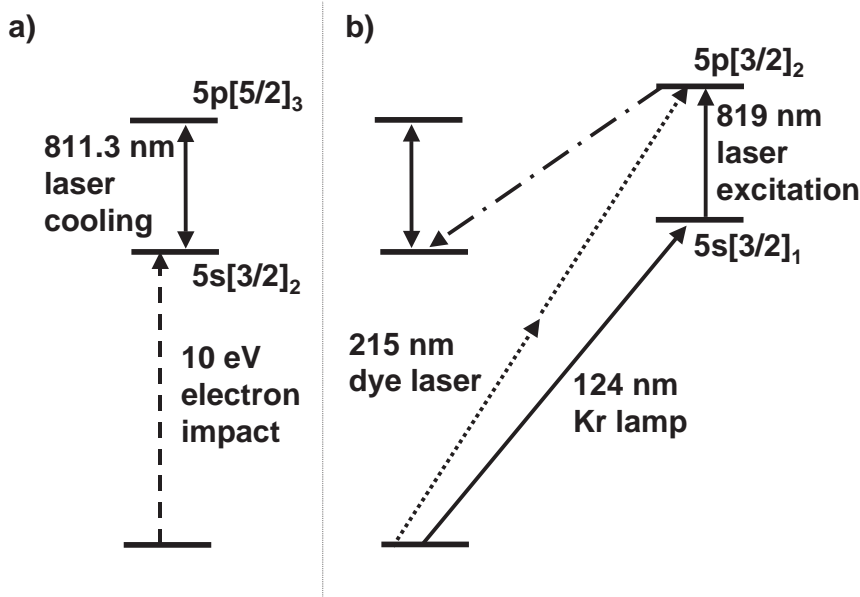, width=12cm}
\end{center}
\caption{The atomic levels of Kr. Figure a) shows the energy
levels relevant to atomic trapping of Kr*, and figure b) shows the
energy levels relevant to resonant and non-resonant optical
excitation of Kr*.} \label{fig:Kr}
\end{figure}

In Kr, the metastable state is the $5s[3/2]_2$ state (sometimes
called the $^3P_2$ state), with a lifetime of
$28$\,s\cite{Lefers}. Laser cooling is accomplished by exciting
the atoms to the $5p[5/2]_3$ state, with a transition wavelength
$\lambda = 811.3$\,nm. Fig.~\ref{fig:Kr} shows some important
levels in Kr.

In order to cool and trap Kr, we must produce $5s[3/2]_2$
metastables in the atomic beam source. Ground-state atoms and
atoms in other excited states will be unaffected by the cooling
and trapping lasers. The excitation can be accomplished either by
electron impact, two-photon resonant optical excitation, or
two-photon non-resonant optical excitation.

In the electron impact method (Fig.~\ref{fig:Kr}a), a plasma
discharge is created and maintained using either high-voltage DC
current\cite{discharge} or radio-frequency electromagnetic
fields\cite{Chen}.  Gas atoms passing through the source are
excited by collisions with electrons in the discharge region.
Discharge sources are robust and will work for any atomic species,
but they have low efficiency, with metastable atom fractions of
$10^{-4} - 10^{-3}$.

The two-photon resonant optical excitation method (Fig~\ref{fig:Kr}b)
uses a Kr lamp to excite atoms to the $5s[3/2]_1$ state using light at
$\lambda = 124$\,nm, followed by laser excitation at $\lambda =
819$\,nm to the $5p[3/2]_2$ state, which decays to the $5s[3/2]_2$
metastable level with a $77\%$ probability.  The remainder of the
excited atoms return to the ground state, and can be re-excited.

Because this is a resonant two-photon process, it can, in principle,
be more efficient than electron impact.  For the present work, it
offers the additional advantage of atomic selectivity: only Kr atoms
will be excited, while background gas atoms will be left in the ground
state.  Optical production of Kr$^*$ has been demonstrated in a gas
cell, and is projected to give a metastable fraction of order
$10^{-2}$ in an atomic beam source\cite{Young}.

A third possibility is two-photon non-resonant optical excitation,
in which two 215 nm photons are absorbed to excite atoms directly
to the $5p[3/2]_2$ state, after which they decay with a $77\%$
probability to the $5s[3/2]_2$ metastable level.  Because the rate
of this excitation scales as the square of the laser intensity,
this would best be accomplished with a high-intensity pulsed
laser. To allow the interaction with the largest possible number
of atoms, the laser can be aimed down the axis of the atomic beam.
A pulse repetition rate on the order of 500 Hz should allow each
ground-state atom in the beam to interact with the laser. We
project a metastable fraction of order $10^{-2}$ using
commercially available pulsed dye laser technology.

The two-photon laser excitation method would simplify the
apparatus somewhat, by requiring only a single additional light
source. This simplification and the expected increase in
excitation efficiency over discharge sources must be weighed
against a probable reduction in trap capture efficiency, as atoms
excited by the laser near the end of the Zeeman slower may not be
decelerated sufficiently to allow their capture in the MOT.

Any of these excitation methods, applied to pure samples of Kr, will
produce an atomic beam with a Kr$^*$ flux of $10^{13}$\,s$^{-1}$
or better. These methods will also work to excite Kr$^*$ atoms from
impurities in Ne or Xe samples, though discharge sources will also
produce Ne$^*$ and Xe$^*$ metastables, which complicates the
collisional dynamics in the sample.

\subsection{Single-Atom Detection System}

Typical metastable atom sources, using abundant isotopes, give MOT
capture rates on the order of $10^7 - 10^8$\,s$^{-1}$, with a
lifetime in the MOT of $\sim 1$\,s. When working with either rare
isotopes or impurities at abundances of $10^{-14}$, the capture
rate drops to $10^{-3}$\,s$^{-1}$ or less, so the average number
of trapped atoms will be less than one. In order to determine the
abundance of rare isotopes we must have a detection system capable
of unambiguously detecting a single trapped atom.

Single-atom detection relies on the fact that atoms trapped in a
MOT scatter photons isotropically at a rate of $10^7$\,s$^{-1}$. A
fraction of this light ($\sim 1\%$) can be collected using lenses,
and detected with an avalanche photodiode (APD). Typical APD
photon counting efficiencies are $\sim 25\%$, giving photon count
rates of $10^3-10^4$\,s$^{-1}$. Contamination by background light
is minimized by focussing the collection system on the very small
trap volume ($< 1$\,mm$^3$), allowing the fluorescence of a single
trapped atom to be easily distinguished. Single-atom detection has
been demonstrated for Cs\cite{Hu} and Kr$^*$\cite{Kr}, with count
rates of $10^4$\,s$^{-1}$ above background and S/N$\sim 40$.

\subsection{Laser System}

In order to place a limit on Kr impurities in other rare gases, it
is sufficient to measure the abundance of $^{84}$Kr, the most
abundant stable isotope ($57\%$ in atmospheric samples).
$^{84}$Kr, like all even atomic mass isotopes of rare gases, has
zero nuclear spin, and thus no hyperfine structure. As a result,
$^{84}$Kr can be cooled and trapped with a single laser frequency,
with no repumping lasers required. The total laser power required
for trapping and cooling is of order $100$\,mW, and can be
obtained using grating-stabilized diode lasers\cite{Lefers}.

\subsection{Collisions and Background}

Inter-species collisions are an important potential complication
for measurements of Kr impurities in Ne or Xe samples. There are
two reactions of particular importance:
\begin{eqnarray}
\rm Ne^* + Kr & \rightarrow & \rm Ne + Kr^+ \label{eq-NeKr}\\
\rm Kr^* + Xe & \rightarrow & \rm Kr + Xe^* \label{eq-KrXe}
\end{eqnarray}
where Ne$^*$ represents the $3s[3/2]_2$ metastable state, and
Xe$^*$ refers to any one of several possible excited states in Xe.
The internal energy of an atom in a metastable state can be
transferred to a ground-state atom during a collision. These
energy transfer reactions can change the excitation efficiency in
the source, and thus the overall loading rate for Kr$^*$.

Eq.~\ref{eq-NeKr} describes Penning ionization of ground-state
krypton in collisions with neon metastables, which have an
internal energy ($16.6$\,eV) greater than the ionization potential
for krypton ($14.0$\,eV). Room temperature cross sections for this
reaction are of order $10^{-15}$\,cm$^2$\cite{Neynaber}, giving a
mean free path for Kr$^*$ in Ne of several meters for the
discharge parameters of Ref.~\cite{Chen}. Moreover, this reaction
can be entirely avoided by using an optical excitation source,
where only Kr atoms are excited. The internal energy of Kr$^*$
($9.9$\,eV) is too low to excite any state in ground-state neon,
so the effect of collisions between excited Kr and ground-state Ne
should be negligible.

The Kr$^*$-Xe collisions described by Eq.~\ref{eq-KrXe} are more
problematic.  In this case, the internal energy of the Kr$^*$
metastable is transferred to a ground-state Xe atom, populating
high-lying excited states in Xe, which decay rapidly.  As energy
is transferred from Kr to Xe, this reaction cannot be completely
avoided by optical excitation.

Room temperature cross sections for these Kr$^*$-Xe collisions are
of order $10^{-14}$\,cm$^2$\cite{Gedanken,Bochkova}, giving a mean
free path for Kr* in Xe under the conditions of Ref.~\cite{Chen}
of $\sim 1$\,cm. This will produce a significant reduction in the
metastable production efficiency. This problem may be alleviated
somewhat by designing the source so that metastables are created
in a region of very low gas density, but collisional quenching in
the source will be a major factor limiting the sensitivity of ATTA
measurements in Xe.

\subsection{Sample Contamination}

Outgassing of Kr atoms from the walls of the vacuum chamber could
in principle limit the accuracy of ATTA for this application. The
fact that the atoms must be in the metastable state in order to be
trapped and detected means that only outgassing in the source
region, where the atoms have a chance of being excited to the
metastable state, needs to be considered. Outgassing in the MOT
chamber will consist of ground-state atoms, which will not be
trapped, and thus will not contribute to the abundance
measurement.

While relatively little is known about noble gas outgassing from
metals at room temperature, both solubility and diffusion for
noble gases are near zero in metals. Hence outgassing is dominated
by nitrogen, carbon monoxide, hydrogen, water, and hydrocarbons.
Some experiments measuring Kr isotope ratios in
meteorites\cite{Lew93} require low Kr outgassing in noble gas mass
spectrometers, which are mostly metal with some attached glassware
and/or ceramic pieces. The standard final cleanup of these systems
involves bake out under vacuum of the complete assembly at a
temperature of 350 C. The background is dominated by the
contributions from the heated parts during the preparation of the
samples, the operation of the valves, and the non-metallic parts
of the system\cite{Lew}.  A $\rm^{82}Kr$ outgassing rate of $\rm 5
\times 10^{-15}$ cc STP is typical for outgassing times of the
order of 60 minutes and chamber areas of the order of 1000 cm.

Based on this outgassing rate, we can expect $\rm 2 \times 10^{1}$
$\rm ^{84}Kr$ atoms/s outgassed into a source chamber of area $100\,
\rm cm^{2}$\cite{Chen}. Given a sample injection rate of $7 \times
10^{16}$ atoms/s, this gives a Kr background of $\rm 3 \times
10^{-16}$ Kr/Ne or Kr/Xe, assuming that Kr atoms from the walls
are excited in the source and captured in the trap with equal
efficiency as Kr atoms from the sample.  In a future ATTA setup
for measuring Kr/Ne or Kr/Xe, further improvements in Kr
outgassing are likely, assuming sustained baking of the vacuum
chamber and careful choice of materials.

An important factor that may complicate this analysis is the
embedding of $\rm Kr^{+}$ ions in the walls during the operation
of a discharge source. These atoms can emerge from the walls
later, either through normal outgassing, or by being driven out by
later ion impacts. Du \textit{et al.}\cite{Du} report this as a
source of cross-sample contamination in experiments to measure
isotopic abundances in atmospheric samples, accounting for more
than 1\% of their Kr signal. This ``memory effect'' can be
minimized by aligning and testing the trap system with pure argon
(the laser cooling wavelengths for Ar$^*$ and Kr$^*$ differ by
less than $1\,$nm, and thus use the same lasers and optics). Even
so, it is potentially a serious problem for a system using a
discharge source when switching between reference samples with
relatively high Kr abundance, and the ultra-pure samples of
interest. The effect can also be avoided by using an optical
excitation source, which will not produce ions or embed atoms in
the source walls.

\section{Discussion}

The ATTA method should allow highly sensitive detection of
Kr atoms in Ne or Xe gases.  Lu \textit{et al.} claim a capture
efficiency in their trap of about 1 part in $10^{7}$, fairly
typical for beam loading systems\cite{Kr}.  Their basic atom
source consumes $7 \times 10^{16}$ atoms per second, which would
give a trap loading rate of about $4 \times 10^{9}$ of the most
abundant isotope ($\rm^{84}Kr$, 57\%).  Making the simple
approximation that the Kr capture efficiency would be the same in
a different gas, one gets a sensitivity of $3 \times 10^{-14}$ in
3 hours of integration.  This assumes a discharge source with
$10^{-4}$ metastable fraction, as was used in \cite{Kr}.  If
optical excitation is used, one could achieve a significantly
higher metastable fraction, potentially enhancing the Kr capture
efficiency and loading rate.

Based on the above simple calculations, we estimate that trace
quantities of Kr in Ne and Xe gases could be detected using Atom
Trap Trace Analysis. Sensitivities as high as 1 part in $3 \times
10^{14}$ should be possible using a simple discharge source, and
even higher sensitivities should be possible using optical
excitation of the Kr. Similar sensitivities should be possible for
measuring trace impurities of Ar.  This technique may prove
valuable for evaluating radioactive contamination in experiments
using condensed noble gases for the detection of low energy
neutrinos or dark matter.

\section{Acknowledgements}

We thank Dr.\,Zheng-Tian Lu, Dr.\,Peter Mueller, and Dr.\,Herman
Beijerinck for valuable discussions.
C.O. is supported by a grant from the Research Corporation.

\end{document}